\documentclass[twocolumn,english]{article}
\usepackage[T1]{fontenc}
\usepackage{amssymb}
\usepackage{graphicx}

\makeatletter
\usepackage{a4}
\input{epsfig.sty}
\def\fnum@table{\tablename~{\bf\thetable}}
\def\fnum@figure{\figurename~{\bf\thefigure}}
\def\tablename{\footnotesize{\bf Table}}
\def\figurename{\footnotesize{\bf Figure}}

\def\be{\begin{equation}}
\def\ee{\end{equation}}

\makeatother

\usepackage{babel}
\begin{document}

\title{\textbf{Double parton scattering: Impact of nonperturbative
parton correlations}}

\author{Sergey Ostapchenko$^{1,2}$ and Marcus Bleicher$^{1,3}$\\
$^1$\textit{\small Frankfurt Institute for Advanced Studies, 
 60438 Frankfurt am Main, Germany}\\
$^2$\textit{\small D.V. Skobeltsyn Institute of Nuclear Physics,
Moscow State University, 119992 Moscow, Russia }\\
$^3$\textit{\small Institute for Theoretical Physics, Goethe-Universitat,
 60438 Frankfurt am Main, Germany}
}

\maketitle
\begin{center}
\textbf{Abstract}
\par\end{center}

We apply the phenomenological Reggeon field theory framework to investigate
the relative importance of perturbative and nonperturbative multiparton
correlations for the treatment of double parton scattering (DPS) in
proton-proton collisions. We obtain a significant correction to the
so-called effective cross section for DPS due to nonperturbative parton
splitting. When combined with the corresponding perturbative contribution,
this results in a rather weak energy and transverse momentum dependence
of the effective cross section, in agreement with experimental observations
at the Tevatron and the Large Hadron Collider.
In addition, we observe that color fluctuations have a sizable
impact on the calculated rate of double parton scattering   and on
 the relative importance of the perturbative parton splitting
mechanism.

\section{Introduction\label{intro.sec}}

There is currently  considerable interest, both from the theoretical
and the experimental sides, to study multiparton interactions. Apart
from being an efficient tool for probing the spatial distribution of
partons in hadrons, such processes constitute a major background for searches
of new physics. Consequently,
the  treatment of multiparton interactions
is an important ingredient of contemporary Monte Carlo generators
of hadronic collisions, both at the energies of the
LHC and at yet higher cosmic ray energies.

Indirect evidence for multiple inelastic parton scatterings comes 
from a comparison of calculated inclusive cross sections for (mini)jet
production in proton-proton collisions with the total inelastic $pp$
cross section $\sigma ^{\rm inel}_{pp}$: for sufficiently small jet 
transverse momenta, the
inclusive jet cross section substantially exceeds 
$\sigma ^{\rm inel}_{pp}$ in the very high energy limit. This means
 that a number of jet production processes should take
place simultaneously. Direct evidence for  double parton interactions
has been obtained experimentally, both at the Tevatron and LHC 
\cite{abe97,aba10,aad13,cha14}.

The pioneering theoretical works on multiparton interactions (MPIs)
date 30 years back \cite{pav82,she82,mek85}, together with the 
first attempts to implement
them in Monte Carlo generators \cite{sjo87}. Recently,
considerable   progress has been achieved in understanding the
MPI physics and developing the theory for multiparton interactions
\cite{blo11,blo12,blo14,rys11,die12,gau13,blo13,sal14}.

The relative rate of double parton scattering (DPS) may be quantified
by the so-called effective cross section defined (here, for the case
of two hadronic dijets production in proton-proton collisions) as
\begin{equation}
\sigma_{pp}^{{\rm eff}}(s,p_{{\rm t}}^{{\rm cut}})=\frac{1}{2}\,\frac{\left[\sigma_{pp}^{2{\rm jet}}(s,p_{{\rm t}}^{{\rm cut}})\right]^{2}}{\sigma_{pp}^{4{\rm jet(DPS)}}(s,p_{{\rm t}}^{{\rm cut}})}\,,\label{eq:sigeff}
\end{equation}
where $\sigma_{pp}^{2{\rm jet}}(s,p_{{\rm t}}^{{\rm cut}})$ is the
inclusive cross section for the production of a pair of jets of transverse
momentum $p_{{\rm t}}>p_{{\rm t}}^{{\rm cut}}$, $\sigma_{pp}^{4{\rm jet(DPS)}}(s,p_{{\rm t}}^{{\rm cut}})$
is the cross section for DPS production of two dijets of $p_{{\rm t}}>p_{{\rm t}}^{{\rm cut}}$,
 (1/2) is a symmetry factor, and $s$ is the center-of-mass energy squared.

The inclusive jet cross section is defined by the usual collinear
factorization ansatz
\begin{eqnarray}
\sigma_{pp}^{2{\rm jet}}(s,p_{{\rm t}}^{{\rm cut}})=\int\! dx^{+}dx^{-}\int_{p_{{\rm t}}>p_{{\rm t}}^{{\rm cut}}}\! dp_{{\rm t}}^{2}\nonumber \\
\times\sum_{I,J=q,\bar{q},g}f_{I}(x^{+},M_{{\rm F}}^{2})\, f_{J}(x^{-},M_{{\rm F}}^{2})\,\frac{d\sigma_{IJ}^{2\rightarrow2}}{dp_{{\rm t}}^{2}}\,,\label{eq:sig-2jet}
\end{eqnarray}
with $x^{\pm}$ being the light-cone momentum fractions, 
$d\sigma_{IJ}^{2\rightarrow2}/dp_{{\rm t}}^{2}$   the parton
scatter cross section, and $f_{I}(x,M_{{\rm F}}^{2})$ the parton $I$
momentum distribution function (PDF) evaluated at the factorization
scale $M_{{\rm F}}^{2}$.

In turn, $\sigma_{pp}^{4{\rm jet(DPS)}}$ involves the so-called generalized
two-parton distribution ($_{2}{\rm GPD}$) $F_{I_{1}I_{2}}^{(2)}(x_{1},x_{2},q_{1}^{2},q_{2}^{2},\Delta\! b)$
which describes the momentum distribution of a pair of partons probed at
the virtuality scales $q_{1}^{2}$ and $q_{2}^{2}$,   separated
by the transverse distance $\Delta\! b$  \cite{blo11,blo12,gau13}:%
\footnote{Here and in the following we use $_{2}{\rm GPDs}$ in the impact parameter
space, which are related to the ones introduced in
Refs.\ \cite{blo11,blo12} via a
Fourier transform.%
}
\begin{eqnarray}
\sigma_{pp}^{4{\rm jet(DPS)}}(s,p_{{\rm t}}^{{\rm cut}})=\frac{1}{2}\int\! dx_{1}^{+}dx_{2}^{+}dx_{1}^{-}dx_{2}^{-}\nonumber \\
\times\int_{p_{{\rm t}_{1}}\!,p_{{\rm t}_{2}}>p_{{\rm t}}^{{\rm cut}}}\!\! dp_{{\rm t}_{1}}^{2}\, dp_{{\rm t}_{2}}^{2}\!\sum_{I_{1},I_{2},J_{1},J_{2}}\!\frac{d\sigma_{I_{1}J_{1}}^{2\rightarrow2}}{dp_{{\rm t}_{1}}^{2}}\,\frac{d\sigma_{I_{2}J_{2}}^{2\rightarrow2}}{dp_{{\rm t}_{2}}^{2}}\nonumber \\
\times\int d^{2}\Delta\! b\: F_{I_{1}I_{2}}^{(2)}(x_{1}^{+},x_{2}^{+},M_{{\rm F}_{1}}^{2},M_{{\rm F}_{2}}^{2},\Delta\! b)\nonumber \\
\times F_{J_{1}J_{2}}^{(2)}(x_{1}^{-},x_{2}^{-},M_{{\rm F}_{1}}^{2},M_{{\rm F}_{2}}^{2},\Delta\! b)\,.\label{eq:sig-DPS}
\end{eqnarray}

In the simplest approach, one neglects multiparton correlations
and expresses $_{2}{\rm GPDs}$ as a convolution of generalized parton
distributions (GPDs) for two independent partons:
\begin{eqnarray}
F_{I_{1}I_{2}}^{(2)}(x_{1},x_{2},q_{1}^{2},q_{2}^{2},\Delta\! b)\nonumber \\
=\int\! d^{2}b'\, G_{I_{1}}(x_{1},q_{1}^{2},b')\, G_{I_{2}}(x_{2},q_{2}^{2},|\vec{b'}-\vec{\Delta\! b}|)\,.\label{eq:2GPD-fact}
\end{eqnarray}

Without a loss of generality,
 we may express $G_{I}$ via the usual PDFs $f_{I}(x,q^{2})$ and the parton
distribution in transverse space $\rho_{I}$ as
\begin{equation}
G_{I}(x,q^{2},b)=f_{I}(x,q^{2})\,\rho_{I}(b,x,q^{2})\,,\label{eq:GPD-fact}
\end{equation}
with $\int\! d^{2}b\,\rho_{I}(b,x,q^{2})=1$. One can thus cast the respective
``(2v2)'' contribution to $\sigma_{pp}^{4{\rm jet(DPS)}}$ in the
form \cite{pav82}
\begin{eqnarray}
\sigma_{pp}^{4{\rm jet(2v2)}}(s,p_{{\rm t}}^{{\rm cut}})=\frac{1}{2}\int\! d^{2}b\nonumber \\
\times\left[\int\! dx^{+}dx^{-}\int_{p_{{\rm t}}>p_{{\rm t}}^{{\rm cut}}}\! dp_{{\rm t}}^{2}\,\sum_{I,J}f_{I}(x^{+},M_{{\rm F}}^{2})\right.\nonumber \\
\times\left.f_{J}(x^{-},M_{{\rm F}}^{2})\,\frac{d\sigma_{IJ}^{2\rightarrow2}}{dp_{{\rm t}}^{2}}\,\Omega_{IJ}(x^{+},x^{-},M_{{\rm F}}^{2},b)\right]^{2}.\label{eq:sig-DPS-2v2}
\end{eqnarray}
Here
\begin{eqnarray}
\Omega_{IJ}(x^{+},x^{-},M_{{\rm F}}^{2},b)=\int\! d^{2}b'\,\rho_{I}(b',x^{+},M_{{\rm F}}^{2})\nonumber \\
\times\rho_{J}(|\vec{b}-\vec{b}'|,x^{-},M_{{\rm F}}^{2})
\end{eqnarray}
 are parton transverse overlap functions. In particular, assuming
a simple universal Gaussian distribution for all partons,
\begin{equation}
\rho_{I}(b,x,q^{2})=\tilde{\rho}(b)=\frac{1}{\pi R_{p}^{2}}\, e^{-b^{2}/R_{p}^{2}}\,,\label{eq:rho-uni}
\end{equation}
one obtains for the respective effective cross section $\sigma_{pp}^{{\rm eff(2v2)}}=4\pi R_{p}^{2}$,
i.e.~$\sigma_{pp}^{{\rm eff(2v2)}}$ is related to the effective
area occupied by partons in the proton.

In reality, one has to take into consideration Gribov's transverse
diffusion which produces a larger transverse spread for partons at
smaller $x$ \cite{bro94,fra02}. Moreover, as follows from simple dimensional
considerations and is supported by HERA data \cite{adl00,che02}, the rate of 
the transverse
diffusion is slower for partons of larger virtuality \cite{fra04}. 
Thus, one expects
$\sigma_{pp}^{{\rm eff(2v2)}}$ to increase with $s$  due to the
longer rapidity range available for the parton evolution, 
resulting in a larger
transverse spread, and to decrease with $p_{t}^{{\rm cut}}$,
due to a smaller part of the parton cascade developing in the low-$q^{2}$ region.

Recently, it has been demonstrated that substantial corrections to
this simple picture arise from parton correlations induced by a perturbative
parton splitting \cite{blo11,blo12,blo14,rys11,gau13,sni10}. 
In the latter case, two (say, projectile)
partons participating in the two hard processes are no longer independent
but emerge from the same ``parent'' parton of relatively high virtuality
$|q^{2}|>Q_{0}^{2}\gg\Lambda_{{\rm QCD}}^{2}$ and are close to each
other in the transverse plane. The respective ``${\rm (2v1)}_{{\rm h}}$''
contribution to $\sigma_{pp}^{4{\rm jet(DPS)}}$ can be defined as
\cite{blo11,blo12,blo14,rys11,gau13}
\begin{eqnarray}
\sigma_{pp}^{4{\rm jet(2v1)}_{{\rm h}}}(s,p_{{\rm t}}^{{\rm cut}})=\frac{1}{2}\sum_{L}\int\!\frac{dx}{x^{2}}\int_{q^{2}>Q_{0}^{2}}\!\frac{dq^{2}}{q^{2}}\nonumber \\
\times f_{L}(x,q^{2})\,\frac{\alpha_{{\rm s}}}{2\pi}\sum_{K}\int\!\!\frac{dz}{z(1-z)}\, P_{L\rightarrow K(K')}^{{\rm AP}}(z)\nonumber \\
\times\int\! dx_{1}^{+}dx_{2}^{+}dx_{1}^{-}dx_{2}^{-}\int_{p_{{\rm t}_{1}}\!,p_{{\rm t}_{2}}>p_{{\rm t}}^{{\rm cut}}}\!\! dp_{{\rm t}_{1}}^{2}\, dp_{{\rm t}_{2}}^{2}\!\nonumber \\
\times\sum_{I_{1},I_{2},J_{1},J_{2}}E_{K\rightarrow I_{1}}(\frac{x_{1}^{+}}{zx},q^{2},M_{{\rm F}_{1}}^{2})\nonumber \\
\times E_{K'\rightarrow I_{2}}(\frac{x_{2}^{+}}{(1-z)x},q^{2},M_{{\rm F}_{2}}^{2})\,\frac{d\sigma_{I_{1}J_{1}}^{2\rightarrow2}}{dp_{{\rm t}_{1}}^{2}}\,\frac{d\sigma_{I_{2}J_{2}}^{2\rightarrow2}}{dp_{{\rm t}_{2}}^{2}}\nonumber \\
\times\int\! d^{2}b\, G_{J_{1}}(x_{1}^{-},M_{{\rm F}_{1}}^{2},b)\, G_{J_{2}}(x_{2}^{-},M_{{\rm F}_{2}}^{2},b)\,,\label{eq:sig-DPS-1v2}
\end{eqnarray}
where $P_{I\rightarrow J(J')}^{{\rm AP}}$ is the Altarelli-Parisi
splitting kernel and $E_{I\rightarrow J}(z,q^{2},Q^{2})$ is the solution
of the DGLAP equations with the initial condition $E_{I\rightarrow J}(z,q^{2},q^{2})=\delta_{I}^{J}\,\delta(1-z)$,
which describes parton evolution from the scale $q^{2}$ to $Q^{2}$.
Despite being suppressed by an additional power of $\alpha_{{\rm s}}$,
this contribution receives strong collinear enhancements \cite{blo11,blo12}
 and, for
sufficiently high $p_{{\rm t}}^{{\rm cut}}$ and $Q_{0}\sim1$ GeV,
appears to be comparable to $\sigma_{pp}^{4{\rm jet(2v2)}}$, thus
reducing the effective cross section {[}Eq.~(\ref{eq:sigeff}){]}
by a factor of  2  \cite{blo14,gau14}.   Implementing the mechanism in
the PYTHIA Monte Carlo generator allows one to obtain a consistent
description of multiparton interactions both for high $p_t$ jet production
and in underlying events \cite{blo15}.

For decreasing $p_{{\rm t}}^{{\rm cut}}$, the contribution of the
perturbative parton splitting goes down due to the reduced kinematic
space for the parton evolution. However, as suggested in
\cite{blo14,blo12a},
additional important corrections to the DPS cross section should arise
from nonperturbative parton correlations, e.g.~ones related to nonperturbative
parton splitting at $|q^{2}|<Q_{0}^{2}$. It is the goal of the present
work to estimate the magnitude of such corrections, using a phenomenological
Reggeon field theory (RFT) \cite{gri68} approach. In Section \ref{sec:DPS-PPP},
we describe the treatment of double parton scattering in the enhanced
Pomeron framework, as implemented in the QGSJET-II model \cite{ost06a,ost11}.
 In Section
\ref{sec:results}, we present some numerical results and discuss
the obtained energy and $p_{t}$ dependence of the DPS cross sections.
Finally, we conclude in Section \ref{sec:Outlook}.

\section{Double parton scattering in the
enhanced Pomeron framework\label{sec:DPS-PPP}}

We are going to investigate DPS using the enhanced Pomeron framework
\cite{kan73,kai86,ost06}, as implemented 
in the QGSJET-II model \cite{ost06a,ost11}.
The approach takes into consideration the contributions of Pomeron-Pomeron
interaction (so-called enhanced) diagrams. At the parton level, such graphs
describe the rescattering of intermediate partons in the parton cascades
off the parent hadrons and off each other. 

To obtain a coherent treatment of both ``soft'' and ``hard'' interaction
processes, general parton cascades are split into two parts, as described
in more detail in Refs.\  \cite{dre99,dre01,ost02}. The hard part is characterized
by high enough parton virtualities $|q^{2}|>Q_{0}^{2}$, $Q_{0}$
being some cutoff for pQCD being applicable, and is treated by means
of the DGLAP evolution equations. In turn, the nonperturbative
soft part involves low-$q^2$ ($|q^{2}|<Q_{0}^{2}$) partons and is described
by a phenomenological soft Pomeron asymptotic.

To treat low mass diffraction and the related absorptive (inelastic
screening) effects, one employs a Good-Walker-type \cite{goo60} framework,
considering the interacting protons to be represented by a superposition
of a number of eigenstates which diagonalize the scattering matrix,
characterized by different couplings to Pomerons \cite{kai79}. The
respective partonic interpretation is based on the color fluctuations
picture \cite{fra08}, i.e. the representation of the proton wave
function by a superposition of parton Fock states of different sizes.
Fock states of larger transverse size are characterized by lower (more
dilute) spatial parton densities, while more compact ones are more
densely packed with partons. As will be demonstrated in the following,
such color fluctuations have important consequences both for the total
rate of double parton scattering and for the relative importance
 of the different DPS contributions.

\begin{figure}[tbh]
\begin{centering}
\includegraphics[width=4cm,height=6cm]{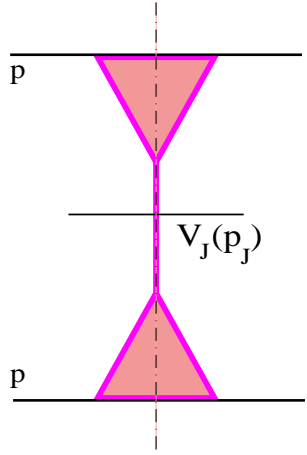}\caption{Schematic view for the general RFT diagram for inclusive jet production
in $pp$ collisions: the projectile and target  triangles  consist
of  fanlike enhanced Pomeron graphs; $V_{J}(p_{J})$ is the parton
$J$ emission vertex from a cut Pomeron. The cut plane is shown by
the vertical dotted-dashed line.\label{fig:2jet}}

\par\end{centering}

\end{figure}

\begin{figure*}[t]
\centering{}\includegraphics[width=12cm,height=2.2cm]{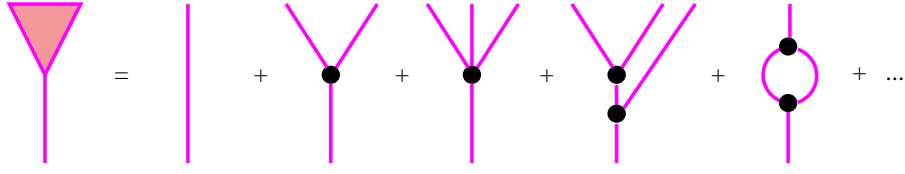}\caption{Examples of enhanced Pomeron graphs of lowest orders, contributing
to the structure of the projectile and target triangles in Fig.~\ref{fig:2jet};
Pomerons are shown by thick lines and multi-Pomeron vertices by filled
circles.\label{fig:triangle}}
\end{figure*}

Let us start with the inclusive cross section for high
transverse momentum  ($p_{{\rm t}}>p_{{\rm t}}^{{\rm cut}}$)
jet production, which is described by Kancheli-Mueller-type 
diagrams depicted in Fig.~\ref{fig:2jet}.
 The internal structure of the projectile and target ``triangles''
of Fig.~\ref{fig:2jet} is explained in Fig.~\ref{fig:triangle}.
 The basic contribution is a single Pomeron emission by the parent
hadron {[}1st graph in the  rhs  of  
Fig.~\ref{fig:triangle}{]},
which corresponds to an ``elementary'' parton cascade. Absorptive
corrections to that process involve Pomeron-Pomeron interactions and
arise from the rescattering of intermediate partons in the cascade off
the parent hadron (2nd, 3rd, and 4th graphs in the rhs) and off each
other (5th graph in the rhs).

Neglecting absorptive corrections due to enhanced Pomeron diagrams,
the inclusive cross section
 $\sigma_{pp}^{2{\rm jet(no}\,{\rm abs)}}(s,p_{{\rm t}}^{{\rm cut}})$
for the production of a pair of jets of transverse momentum $p_{{\rm t}}>p_{{\rm t}}^{{\rm cut}}$
is defined as \cite{dre01,ost02}
\begin{eqnarray}
\sigma_{pp}^{2{\rm jet(no}\,{\rm abs)}}(s,p_{{\rm t}}^{{\rm cut}})=\sum_{i,j}C_{i}\, C_{j}\int\! d^{2}b'\, d^{2}b''\nonumber \\
\times\int\!\frac{dx^{+}}{x^{+}}\frac{dx^{-}}{x^{-}}\,\sum_{I,J}\chi_{(i)I}^{\mathbb{P}_{{\rm soft}}}(s_{0}/x^{+},b')\nonumber \\
\times\chi_{(j)J}^{\mathbb{P}_{{\rm soft}}}(s_{0}/x^{-},b'')\,\sigma_{IJ}^{{\rm QCD}}(x^{+}x^{-}s,Q_{0}^{2},p_{{\rm t}}^{{\rm cut}})\,,\label{eq:sigma-2jet-noabs}
\end{eqnarray}
where
\begin{eqnarray}
\sigma_{IJ}^{{\rm QCD}}(x^{+}x^{-}s,Q_{0}^{2},p_{{\rm t}}^{{\rm cut}})=\int_{p_{{\rm t}}>p_{{\rm t}}^{{\rm cut}}}\! dp_{{\rm t}}^{2}\nonumber \\
\times\int\! dz^{+}dz^{-}\,\sum_{I',J'}\frac{d\sigma_{I'J'}^{2\rightarrow2}(x^{+}x^{-}z^{+}z^{-}s,p_{t}^{2})}{dp_{{\rm t}}^{2}}\nonumber \\
\times E_{I\rightarrow I'}(z^{+},Q_{0}^{2},M_{{\rm F}}^{2})\, 
E_{J\rightarrow J'}(z^{-},Q_{0}^{2},M_{{\rm F}}^{2})\label{eq:sig-lad}
\end{eqnarray}
is the contribution of the DGLAP ladder, corresponding to the production
of a pair of jets of $p_{{\rm t}}>p_{{\rm t}}^{{\rm cut}}$, with
ladder leg partons of types $I,J$ {[}sea (anti)quarks or gluons%
\footnote{For brevity, we shall not discuss explicitly valence quark
contributions,
which are of secondary importance for the present analysis.%
}{]}, characterized by the virtuality $Q_{0}^{2}$ and fractions $x^{+},x^{-}$
of the parent hadrons' light-cone momenta. In turn, $\chi_{(i)I}^{\mathbb{P}_{{\rm soft}}}$
is the eikonal describing a soft Pomeron exchange between the parent
proton represented by its diffractive eigenstate $|i\rangle$ and
the ladder leg parton $I$;
\begin{equation}
\chi_{(i)I}^{\mathbb{P}_{{\rm soft}}}(\hat{s},b)\propto\frac{\lambda_{i}\,\hat{s}^{\alpha_{\mathbb{P}}-1}}{R_{i}^{2}+\alpha'_{\mathbb{P}}\,\ln\hat{s}}\, e^{-\frac{b^{2}}{4(R_{i}^{2}+\alpha'_{\mathbb{P}}\,\ln\hat{s})}}\label{eq:Psoft}
\end{equation}
for sufficiently large $\hat{s}$; $\alpha{}_{\mathbb{P}}$ and $\alpha'_{\mathbb{P}}$
are, respectively, the intercept and the slope of the soft Pomeron Regge
trajectory. $C_{i}$ is the partial weight of Fock state $|i\rangle$
while $\lambda_{i}$ and $R_{i}^{2}$ are the relative strength and
the slope for Pomeron coupling to $|i\rangle$, $\sum_{i}C_{i}\,\lambda_{i}=1$.

Including absorptive corrections due to enhanced Pomeron graphs but
neglecting for simplicity contributions of Pomeron loop diagrams (of
the kind of the 5th graph in the rhs of Fig.~\ref{fig:triangle}),
which prove not to be 
essential in the kinematic range studied in this
work, the projectile and target  triangles  in Fig.~\ref{fig:2jet}
consist of ``fanlike''  Pomeron graphs (examples of contributions
of lowest orders are 1st, 2nd, 3rd, and 4th graphs in the rhs of Fig.~\ref{fig:triangle}).
Thus, Eq.~(\ref{eq:sigma-2jet-noabs}) transforms to \cite{ost06a}
\begin{eqnarray}
\sigma_{pp}^{2{\rm jet}}(s,p_{{\rm t}}^{{\rm cut}})=\sum_{i,j}C_{i}\, C_{j}\int\! d^{2}b'\, d^{2}b''\nonumber \\
\times\int\! dx^{+}\, dx^{-}\,\sum_{I,J}\tilde{f}_{I}^{(i)}(x^{+},b')\,\tilde{f}_{J}^{(j)}(x^{-},b'')\nonumber \\
\times\sigma_{IJ}^{{\rm QCD}}(x^{+}x^{-}s,Q_{0}^{2},p_{{\rm t}}^{{\rm cut}})\,,\label{eq:sigma-2jet-abs}
\end{eqnarray}
with
\begin{eqnarray}
x\,\tilde{f}_{I}^{(i)}(x,b)=\chi_{(i)I}^{\mathbb{P}_{{\rm soft}}}(s_{0}/x,b)+G\int\! d^{2}b'\int\!\frac{dx'}{x'}\nonumber \\
\times\left[1-e^{-\chi_{(i)}^{{\rm fan}}(s_{0}/x',b')}-\chi_{(i)}^{{\rm fan}}(s_{0}/x',b')\right]\nonumber \\
\times\chi_{\mathbb{P}I}^{\mathbb{P}_{{\rm soft}}}(s_{0}\, x'/x,|\vec{b}-\vec{b}'|)\,.\label{eq:GPD-Q0}
\end{eqnarray}
Here $\chi_{(i)}^{{\rm fan}}$ is the solution of the fan diagram
equation 
\begin{eqnarray}
\chi_{(i)}^{{\rm fan}}(\hat{s},b)=\chi_{(i)\mathbb{P}}^{\mathbb{P}_{{\rm soft}}}(\hat{s},b)+G\int\! d^{2}b'\int\!\frac{dx'}{x'}\nonumber \\
\times\left[1-e^{-\chi_{(i)}^{{\rm fan}}(s_{0}/x',b')}-\chi_{(i)}^{{\rm fan}}(s_{0}/x',b')\right]\nonumber \\
\times\chi_{\mathbb{PP}}^{\mathbb{P}_{{\rm soft}}}(x'\hat{s},|\vec{b}-\vec{b}'|)\,,\label{eq:fan}
\end{eqnarray}
where the eikonal $\chi_{(i)\mathbb{P}}^{\mathbb{P}_{{\rm soft}}}(\hat{s},b)$
corresponds to soft Pomeron exchange between proton's diffractive
eigenstate $|i\rangle$ and a multi-Pomeron vertex, which depends
on $\hat{s}$, $b$, $\lambda_{i}$, $R_{i}^{2}$ like $\chi_{(i)I}^{\mathbb{P}_{{\rm soft}}}$
in Eq.~(\ref{eq:Psoft}). $\chi_{\mathbb{PP}}^{\mathbb{P}_{{\rm soft}}}$
describes soft Pomeron exchange between two multi-Pomeron vertices,
\begin{equation}
\chi_{\mathbb{PP}}^{\mathbb{P}_{{\rm soft}}}(\hat{s},b)=
\frac{\gamma_{\mathbb{P}}^{2}\,\hat{s}^{\alpha_{\mathbb{P}}-1}}
{\alpha'_{\mathbb{P}}\,\ln\hat{s}}\, e^{-\frac{b^{2}}{4\alpha'_{\mathbb{P}}\,
\ln\hat{s}}}\,.\label{eq:P-int}
\end{equation}
The eikonal $\chi_{\mathbb{P}I}^{\mathbb{P}_{{\rm soft}}}$, corresponding
to a Pomeron exchanged between a multi-Pomeron vertex and parton $I$,
is obtained from $\chi_{\mathbb{PP}}^{\mathbb{P}_{{\rm soft}}}$ replacing
one factor $\gamma_{\mathbb{P}}$ by a Pomeron-parton vertex, as discussed
in more detail in Ref.\  \cite{ost06a}. 
It is noteworthy that Eqs.~(\ref{eq:sigma-2jet-abs}-\ref{eq:fan})
have been derived in Ref.\  \cite{ost06a}  neglecting
parton transverse diffusion during the perturbative ($|q^{2}|>Q_{0}^{2}$)
evolution and assuming the vertices for the transition
of $m$ into $n$ Pomerons to be of the form  \cite{kai86}
\begin{equation}
G^{(m,n)}=G\,\gamma_{\mathbb{P}}^{m+n}.\label{eq:Gmn}
\end{equation}
 $G$ is related to the triple-Pomeron vertex $r_{3\mathbb{P}}$
as $G=r_{3\mathbb{P}}/(4\pi\gamma_{\mathbb{P}}^{3}$).

Comparing now Eqs.~(\ref{eq:sigma-2jet-abs}) and (\ref{eq:sig-lad})
with Eqs.~(\ref{eq:sig-2jet}) and (\ref{eq:GPD-fact}), we may interpret
$\tilde{f}_{I}^{(i)}(x,b)$ as GPDs $G_{I}^{(i)}(x,Q_{0}^{2},b)$
of sea (anti)quarks and gluons at the virtuality scale $Q_{0}^{2}$
for the Fock state $|i\rangle$  \cite{ost06a}. Let us briefly discuss
their $x$ and $b$ dependence. For sufficiently large $b$, absorptive
corrections are weak and the respective low $x$ behavior of $\tilde{f}_{I}^{(i)}$
is  described by the soft Pomeron asymptotic, $\tilde{f}_{I}^{(i)}(x,b)\propto x^{-\alpha_{\mathbb{P}}}$
for $x\rightarrow0$ {[}cf.~Eqs.~(\ref{eq:GPD-Q0}) and (\ref{eq:Psoft}){]}.
For decreasing $b$, absorptive corrections become stronger, producing
parton shadowing effects. At sufficiently small $x$, one approaches
the flat $x\,\tilde{f}_{I}^{(i)}(x,b)\simeq {\rm const}$ behavior for
$b\simeq 0$ \cite{ost05}.  Let us turn now to the parton spatial distribution.
Because of the transverse diffusion during the soft ($|q^{2}|<Q_{0}^{2}$)
parton cascade, governed by the soft Pomeron slope $\alpha'_{\mathbb{P}}$,
smaller $x$ partons are distributed over a larger transverse area
{[}see Eq.~(\ref{eq:Psoft}){]} at the scale $Q_{0}^{2}$. As the
Pomeron slope is rather small, the difference between the effective
radii of the parton spatial distributions for different Fock states,  
caused by the different slopes $R_{i}^{2}$, should persist down
to rather small values of $x$, being gradually washed out by the transverse
diffusion. It is noteworthy that absorptive corrections produce stronger
shadowing at smaller impact parameters and for smaller $x$, as already
discussed above, which effectively speeds up the diffusion process.%
\footnote{Such an effect has been discussed previously in Ref.\ \cite{glm08}.%
}

GPDs at a higher scale $Q^{2}>Q_{0}^{2}$ are obtained by evolving
the input ones (as mentioned above, we neglect parton transverse diffusion
at $|q^{2}|>Q_{0}^{2})$:
\begin{eqnarray}
G_{I}^{(i)}(x,Q^{2},b)=\sum_{I'}\int_{x}^{1}\!\frac{dz}{z}\, E_{I'\rightarrow I}(z,Q_{0}^{2},Q^{2})\nonumber \\
\times\tilde{f}_{I'}^{(i)}(x/z,b)\,.\label{eq:GPD-evolv}
\end{eqnarray}
Thus, the available rapidity range $\Delta y=\ln(1/x)$ is shared
between the soft ($|q^{2}|<Q_{0}^{2})$ and the hard parton evolution;
the larger the $Q^{2}$, the shorter the soft part. As a consequence, for
a given $x$, partons of higher $Q^{2}$ are distributed over a smaller
transverse area.

\begin{figure*}[tbh]
\centering{}\includegraphics[width=14cm,height=4cm]{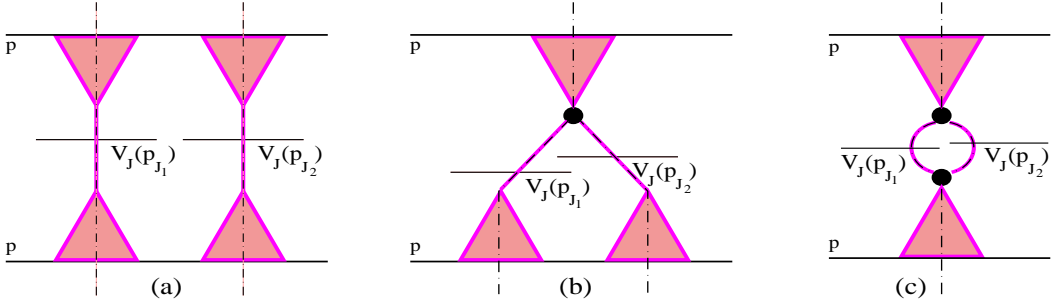}\caption{Schematic view of RFT diagrams for double parton scattering: contribution
of two independent parton cascades (a), projectile parton splitting
(b), and double  parton splitting (c).\label{fig:RFT-double}}
\end{figure*}

Let us now  turn to the DPS contribution to double dijet production.
We start with the graph in Fig.~\ref{fig:RFT-double}(a)
 where the pair of projectile (respectively, target) partons participating
in the two hard processes originates from independent parton cascades.
This leads us to
\begin{eqnarray}
\sigma_{pp}^{4{\rm jet(2v2)}}(s,p_{{\rm t}}^{{\rm cut}})=\frac{1}{2}\sum_{i,j}C_{i}\, C_{j}\int\! d^{2}b\nonumber \\
\times\left[\int\! dx^{+}\, dx^{-}\,\sum_{I,J}\sigma_{IJ}^{{\rm QCD}}(x^{+}x^{-}s,Q_{0}^{2},p_{{\rm t}}^{{\rm cut}})\right.\nonumber \\
\times\left.\int\! d^{2}b'\,\tilde{f}_{I}^{(i)}(x^{+},b')\,\tilde{f}_{J}^{(j)}(x^{-},|\vec{b}-\vec{b}'|)\right]^{2}.\label{eq:sigma-2v2-RFT}
\end{eqnarray}
Interpreting $\tilde{f}_{I}^{(i)}(x,b)$ as GPDs at the scale $Q_{0}^{2}$,
Eq.~(\ref{eq:sigma-2v2-RFT}) is similar to Eq.~(\ref{eq:sig-DPS-2v2})
with one important difference: the pair of projectile (respectively,
target) partons participating in the hard processes is not actually
uncorrelated. Indeed, $\sigma_{pp}^{4{\rm jet(2v2)}}$ in 
Eq.~(\ref{eq:sigma-2v2-RFT})
is obtained by averaging over contributions of different Fock states.
Most importantly, in Fock states with the smallest $R_{i}^{2}$, partons
are more closely packed together, which   enhances the respective
contributions to $\sigma_{pp}^{4{\rm jet(2v2)}}$. On the other hand,
for increasing $s$, the effect is slowly washed out due to the parton
transverse diffusion.

Next, we consider the ``soft splitting'' contribution ``${\rm (2v1)}_{{\rm s}}$''
of Fig.~\ref{fig:RFT-double}(b), for which we obtain (neglecting
Pomeron loop corrections) 
\begin{eqnarray}
\sigma_{pp}^{4{\rm jet(2v1)}_{{\rm s}}}(s,p_{{\rm t}}^{{\rm cut}})=\frac{1}{2}\sum_{i,j}C_{i}\, C_{j}\int\! d^{2}b'\int\!\frac{dx'}{x'}\nonumber \\
\times G\left[1-e^{-\chi_{(i)}^{{\rm fan}}(s_{0}/x',b')}\right]\int\! d^{2}b\left[\int\!\frac{dx^{+}}{x^{+}}\right.\nonumber \\
\times\int\! dx^{-}\int\! d^{2}b''\sum_{I,J}\chi_{\mathbb{P}I}^{\mathbb{P}_{{\rm soft}}}(s_{0}\, x'/x^{+},b'')\nonumber \\
\times\left.\tilde{f}_{J}^{(j)}(x^{-},|\vec{b}-\vec{b}''|)\,\sigma_{IJ}^{{\rm QCD}}(x^{+}x^{-}s,Q_{0}^{2},p_{{\rm t}}^{{\rm cut}})\right]^{2}.\label{eq:1v2s}
\end{eqnarray}
It is noteworthy that the (2v2) {[}Eq.~(\ref{eq:sigma-2v2-RFT}){]}
and the two (projectile and target parton splitting) ${\rm (2v1)}_{{\rm s}}$
{[}Eq.~(\ref{eq:1v2s}){]} contributions can be obtained from Eq.~(\ref{eq:sig-DPS})
if $_{2}{\rm GPDs}$ at the scale $Q_{0}^{2}$ are defined as 
\begin{eqnarray}
F_{I_{1}I_{2}}^{(2)}(x_{1},x_{2},Q_{0}^{2},Q_{0}^{2},\Delta\! b)\nonumber \\
=\sum_{i}C_{i}\int\! d^{2}b'\left\{ \tilde{f}_{_{I_{1}}}^{(i)}(x_{1},b')\,\tilde{f}_{I_{2}}^{(i)}(x_{2},|\vec{b}'-\vec{\Delta\! b}|)\right.\nonumber \\
+\frac{G}{x_{1}x_{2}}\int\!\frac{dx'}{x'}\left[1-e^{-\chi_{(i)}^{{\rm fan}}(s_{0}/x',b')}\right]\int\! d^{2}b''\nonumber \\
\times\left.\chi_{\mathbb{P}I_{1}}^{\mathbb{P}_{{\rm soft}}}(s_{0}x'/x_{1},b'')\,
\chi_{\mathbb{P}I_{2}}^{\mathbb{P}_{{\rm soft}}}(s_{0}x'/x_{2},|\vec{b}''-\vec{\Delta\! b}|)\right\} \label{eq:2GPD-soft}
\end{eqnarray}
and the two branches of the cascade are separately evolved from $Q_{0}^{2}$
to $q_{1}^{2}$, $q_{2}^{2}$:
\begin{eqnarray}
F_{I_{1}I_{2}}^{(2)}(x_{1},x_{2},q_{1}^{2},q_{2}^{2},\Delta\! b)=\sum_{J_{1},J_{2}}\int_{x_{1}}^{1}\!\frac{dz_{1}}{z_{1}}\int_{x_{2}}^{1}\!\frac{dz_{2}}{z_{2}}\nonumber \\
\times E_{J_{1}\rightarrow I_{1}}(z_{1},Q_{0}^{2},q_{1}^{2})\,
 E_{J_{2}\rightarrow I_{2}}(z_{2},Q_{0}^{2},q_{2}^{2})\nonumber \\
\times F_{J_{1}J_{2}}^{(2)}(x_{1}/z_{1},x_{2}/z_{2},Q_{0}^{2},Q_{0}^{2},\Delta\! b)\,.\label{eq:2GPD-evol}
\end{eqnarray}
In addition, such a substitution would generate the loop contribution
of Fig.~\ref{fig:RFT-double}(c), which is neglected in the present
analysis (the respective correction appears to be at the few percent
level).%
\begin{figure*}[tbh]
\centering{}\includegraphics[clip,width=14cm,height=6cm]{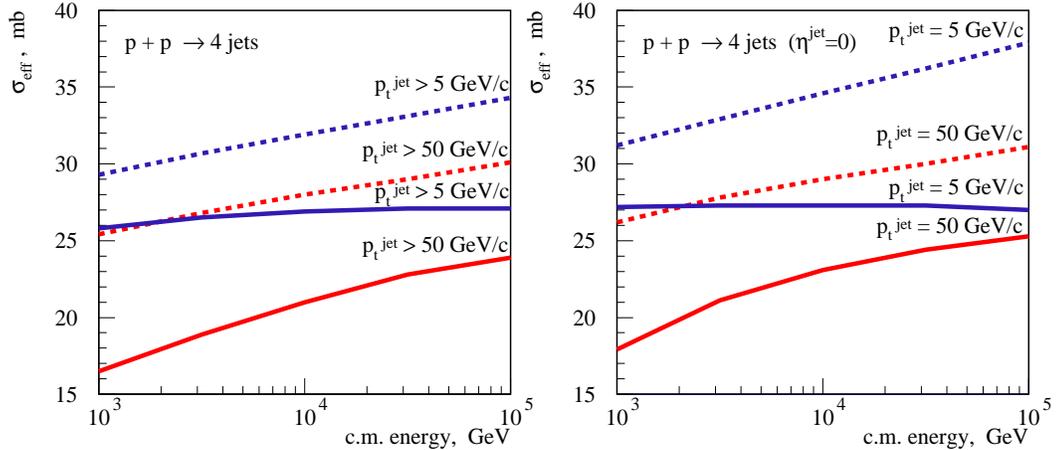}\caption{Energy dependence
 of the effective cross section for the production
of two dijets in double parton scattering. Left: 
 for jet transverse momenta  $p_{{\rm t}}^{{\rm jet}}>5$
GeV/c and $p_{{\rm t}}^{{\rm jet}}>50$ GeV/c (as indicated in the plots),
integrated over the phase space. Right:  for  fixed
 $p_{{\rm t}}^{{\rm jet}}=5$ and 50 GeV/c  and $\eta^{{\rm jet}}=0$.
Solid lines - both the $(2{\rm v}2)$ and ${\rm (2v1)}$ contributions
to DPS taken into account, dashed lines - only the $(2{\rm v}2)$
contribution considered.\label{fig:sigeff-E}}
\end{figure*}%

Finally, we have to add the perturbative parton splitting ${\rm (2v1)}_{{\rm h}}$
contribution, which is not included in the present QGSJET-II model.%
\footnote{The respective RFT treatment was based on phenomenological multi-Pomeron
vertices (\ref{eq:Gmn}), assuming that Pomeron-Pomeron coupling is
dominated by parton processes at low $|q^{2}|<Q_{0}^{2}$.%
} Interpreting $\tilde{f}_{I}^{(i)}(x,b)$ as partial GPDs at the scale
$Q_{0}^{2}$, we obtain, similarly to Eq.~(\ref{eq:sig-DPS-1v2}),
\begin{eqnarray}
\sigma_{pp}^{4{\rm jet(2v1)}_{{\rm h}}}(s,p_{{\rm t}}^{{\rm cut}})=\frac{1}{2}\sum_{i,j}C_{i}\, C_{j}\int_{q^{2}>Q_{0}^{2}}\!\frac{dq^{2}}{q^{2}}\nonumber \\
\times\int\!\frac{dx}{x^{2}}\int\! d^{2}b'\sum_{L}G_{L}^{(i)}(x,q^{2},b')\int\!\frac{dz}{z(1-z)}\,\nonumber \\
\times\frac{\alpha_{{\rm s}}}{2\pi}\sum_{K}P_{L\rightarrow K(K')}^{{\rm AP}}(z)\int\! dx_{1}^{+}dx_{2}^{+}dx_{1}^{-}dx_{2}^{-}\!\nonumber \\
\times\int_{p_{{\rm t}_{1}}\!,p_{{\rm t}_{2}}>p_{{\rm t}}^{{\rm cut}}}\!\! dp_{{\rm t}_{1}}^{2}\, dp_{{\rm t}_{2}}^{2}\sum_{I_{1},I_{2},J_{1},J_{2}}\frac{d\sigma_{I_{1}J_{1}}^{2\rightarrow2}}{dp_{{\rm t}_{1}}^{2}}\,\frac{d\sigma_{I_{2}J_{2}}^{2\rightarrow2}}{dp_{{\rm t}_{2}}^{2}}\nonumber \\
\times E_{K\rightarrow I_{1}}(x_{1}^{+}/x/z,q^{2},M_{{\rm F}_{1}}^{2})\nonumber \\
\times E_{K'\rightarrow I_{2}}(x_{2}^{+}/x/(1-z),q^{2},M_{{\rm F}_{2}}^{2})\nonumber \\
\times\int\! d^{2}b\: G_{J_{1}}^{(j)}(x_{1}^{-},M_{{\rm F}_{1}}^{2},b)\, G_{J_{2}}^{(j)}(x_{2}^{-},M_{{\rm F}_{2}}^{2},b)\,,\label{eq:sig-DPS-1v2-rft}
\end{eqnarray}
where $G_{I}^{(i)}(x,Q^{2},b)$ is defined by Eq.~(\ref{eq:GPD-evolv}).

Because of the smallness of the Pomeron slope $\alpha'_{\mathbb{P}}$,
the soft splitting contribution ${\rm (2v1)}_{{\rm s}}$ of Eq.~(\ref{eq:1v2s})
bears some similarity to the one of perturbative splitting {[}Eq.~(\ref{eq:sig-DPS-1v2-rft}){]}:
the two projectile partons participating in the two hard processes
are close by in the transverse plane {[}the integral over $b''$  in Eq.~(\ref{eq:1v2s})
is dominated by the small $b''\ll b$ region,
cf.~Eq.~(\ref{eq:P-int}){]}. In other words, the second term in
the curly brackets in the rhs of Eq.~(\ref{eq:2GPD-soft}) generates
short range correlations between the two partons in   coordinate
space.

In the next section, we apply Eqs.~(\ref{eq:sigma-2jet-abs}), (\ref{eq:sigma-2v2-RFT}),
(\ref{eq:1v2s}), and (\ref{eq:sig-DPS-1v2-rft}) to investigate the energy
and transverse momentum dependence of the partial DPS cross sections
for the production of two dijets and of the respective effective cross
sections. We shall use the parameter set of the QGSJET-II-04 model \cite{ost11},
which has been obtained by fitting the model to available accelerator
data on the total and elastic proton-proton cross sections, elastic scattering
slope, and total and diffractive structure functions 
$F_{2}$, $F_{2}^{{\rm D}(3)}$.
In particular, we consider two diffractive eigenstates with equal
weights, $C_{1}=C_{2}=1/2$, with the relative strengths $\lambda_{1}=1.6$,
$\lambda_{2}=0.4$, and with the slopes $R_{1}^{2}=2.5\:{\rm GeV}^{-2}$,
$R_{2}^{2}=0.2\:{\rm GeV}^{-2}$; the Pomeron intercept, slope, and
the triple-Pomeron coupling are, respectively, $\alpha{}_{\mathbb{P}}=1.17$,
$\alpha'_{\mathbb{P}}=0.14\:{\rm GeV}^{-2},$ and $r_{3\mathbb{P}}=0.1$
GeV.  For the ``soft-hard'' separation scale, we use $Q_0^2=3$ GeV$^2$  and
the factorization scale in Eq.\ (\ref{eq:sig-lad}) is chosen as
$M_{{\rm F}}^{2}=p_t^2/4$.

\section{Results and discussion\label{sec:results}}

Let us start with the investigation of the   energy dependence of the 
effective  cross section for the production of two hadronic dijets  in double
parton scattering.
In Fig.~\ref{fig:sigeff-E}~(left), 
we plot by solid lines $\sigma_{pp}^{{\rm eff}}$ for the production of jets of
 transverse momenta
 $p_{{\rm t}}^{{\rm jet}}>p_{{\rm t}}^{{\rm cut}}$ for two choices of
 $p_{{\rm t}}^{{\rm cut}}$, integrated over the phase
 space. More specifically,  $\sigma_{pp}^{{\rm eff}}$ is
calculated according to Eq.~(\ref{eq:sigeff}), taking into account
all the discussed contributions to $\sigma_{pp}^{4{\rm jet(DPS)}}$, i.e.
\begin{equation}
\sigma_{pp}^{{\rm eff}}(s,p_{{\rm t}}^{{\rm cut}})=\frac{1}{2}\,
\frac{\left[\sigma_{pp}^{2{\rm jet}}(s,p_{{\rm t}}^{{\rm cut}})\right]^{2}}
{\sum_{\alpha} \sigma_{pp}^{4{\rm jet}(\alpha)}(s,p_{{\rm t}}^{{\rm cut}})}\,,
\label{eq:sigeff-tot}
\end{equation}
where $\alpha={\rm (2v2)}, {\rm (2v1)}_{{\rm s}}, {\rm (2v1)}_{{\rm h}}$.
The corresponding cross sections $\sigma_{pp}^{4{\rm jet(2v2)}}$,
 $\sigma_{pp}^{4{\rm jet}{\rm (2v1)}_{{\rm s}}}$, and
 $\sigma_{pp}^{4{\rm jet}{\rm (2v1)}_{{\rm h}}}$ are defined by 
 Eqs.\ (\ref{eq:sigma-2v2-RFT}), (\ref{eq:1v2s}), and 
 (\ref{eq:sig-DPS-1v2-rft}), respectively, with
 the latter two contributions being multiplied by factor of 2 to account
 for the possibility of parton splitting both on
the projectile and target sides. In addition, we show by dashed lines
 the effective cross section as obtained by considering the simplest (2v2)
DPS configuration only, 
 $\sigma_{pp}^{{\rm eff(2v2)}}=
 \frac{1}{2}\left[\sigma_{pp}^{2{\rm jet}}
 \right]^{2}/\sigma_{pp}^{4{\rm jet(2v2)}}$.
For comparison with earlier studies, we repeat the above calculations for the
case of central ($\eta =0$) production of a pair of dijets of fixed transverse
momentum $p_{{\rm t}}^{{\rm jet}}$, the results
being plotted in  Fig.~\ref{fig:sigeff-E}~(right).%
\begin{figure*}[tbh]
\centering{}\includegraphics[clip,width=14cm,height=6cm]{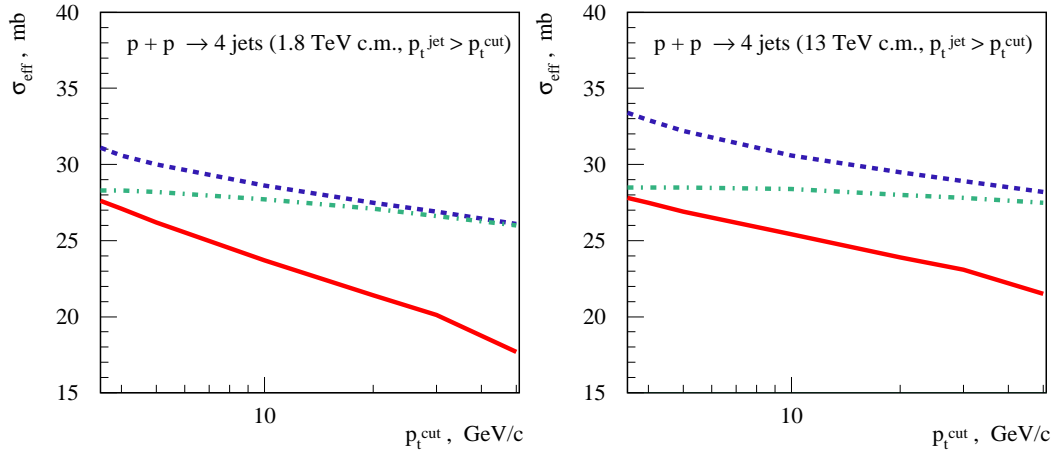}\caption{Effective
 cross section for the production of two dijets of
   $p_{{\rm t}}^{{\rm jet}}>p_{{\rm t}}^{{\rm cut}}$,
integrated over the phase space, in double
parton scattering at $\sqrt{s}=1.8$ TeV (left) and $\sqrt{s}=13$
TeV (right) as a function of $p_{{\rm t}}^{{\rm cut}}$: solid - all
the DPS contributions taken into account, dashed - only the $(2{\rm v}2)$
contribution considered, dotted-dashed - including both the $(2{\rm v}2)$
and ${\rm (2v1)}_{{\rm s}}$ contributions.\label{fig:sigeff-pt}}
\end{figure*}%
 
Looking first at $\sigma_{pp}^{{\rm eff(2v2)}}$  shown by the dashed lines
 in  Fig.~\ref{fig:sigeff-E}, we 
   clearly see the trends discussed
in the Introduction and already observed in previous studies  \cite{blo14,gau14}:
the respective effective cross section increases with $\sqrt{s}$ and decreases
with jet $p_{{\rm t}}$. The energy rise is due to  the
increasing rapidity interval for the parton evolution, hence, also an
extended rapidity range for the soft  ($|q^{2}|<Q_{0}^{2}$) evolution,
which results in a larger parton transverse spread. The decrease of
 $\sigma_{pp}^{{\rm eff}}$ with increasing jet $p_{{\rm t}}$
 for the same $\sqrt{s}$ is caused by a reduction of the soft part 
 of the parton evolution. Indeed, configurations with a larger part 
 of the parton cascade
developing in the hard ($|q^{2}|>Q_{0}^{2}$) regime have stronger
(double logarithmic) enhancement, thus winning over the ones characterized
by a longer soft part. In turn, shorter soft evolution produces a
smaller transverse spread of partons and leads to a smaller effective
cross section.

Interestingly, for sufficiently large $p_{{\rm t}}^{{\rm jet}}$
  the obtained values of $\sigma_{pp}^{{\rm eff(2v2)}}$
are noticeably reduced relative to the results of Refs.\  \cite{blo14,gau14}.
For example,   for central production of two dijets at $\sqrt{s}=1.8$ TeV,
we obtain   $\sigma_{pp}^{{\rm eff(2v2)}}\simeq 27$ mb for
 $p_{{\rm t}}^{{\rm jet}}=50$ GeV/c, to be compared to $\simeq
30-32$ mb in  Refs.\  \cite{blo14,gau14}.
This reduction is caused by averaging over the contributions of 
different Good-Walker Fock states in  Eq.\ (\ref{eq:sigma-2v2-RFT}),
which corresponds to averaging over color fluctuations
in the interacting protons.
 Indeed, the effective cross
section is very sensitive to contributions of parton Fock states of
small transverse size, which are characterized by a higher spatial
parton density and correspondingly by a larger DPS rate, as discussed
previously in Refs.\  \cite{fra08,tre07}.  In other words, the discussed
reduction of  $\sigma_{pp}^{{\rm eff(2v2)}}$ is caused by 
additional correlations
in parton spatial distributions, which are generated by color fluctuations,
compared to the simple factorization ansatz of Eq.\ (\ref{eq:2GPD-fact}).%
\begin{figure*}[tbh]
\raggedright{}\includegraphics[clip,width=14cm,height=6cm]{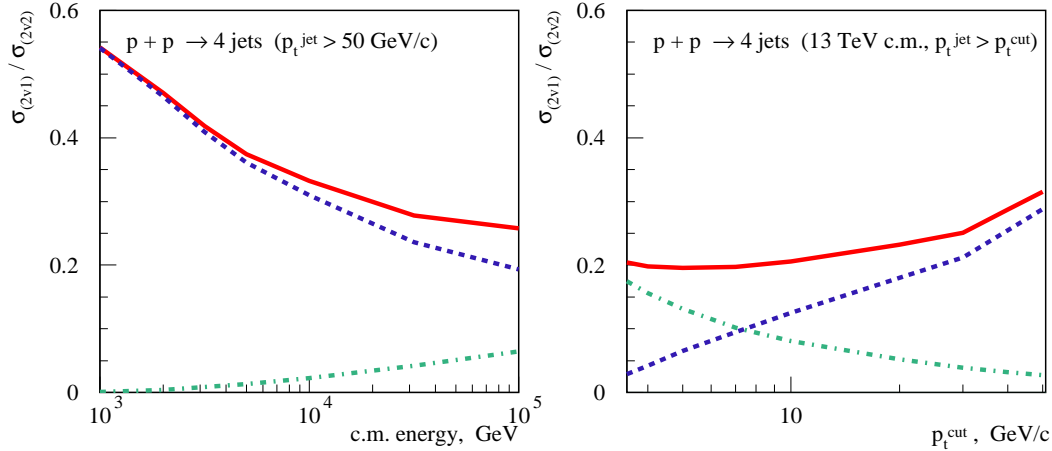}\caption{Energy dependence
 for $p_{{\rm t}}^{{\rm cut}}=50$ GeV/c (left) and
$p_{{\rm t}}^{{\rm cut}}$ dependence for $\sqrt{s}=13$ TeV (right)
of the ratio of the (2v1) to (2v2) contributions to the DPS cross
section for double dijet ($p_{{\rm t}}^{{\rm jet}}>p_{{\rm t}}^{{\rm cut}}$)
production (solid) and partial contributions
 to this ratio from hard
(dashed) and soft (dotted-dashed) parton splitting mechanisms.\label{fig:Reff}}
\end{figure*}%

Let us now turn to $\sigma_{pp}^{{\rm eff}}$ calculated with the soft
and hard parton splittings taken into account, shown by the solid lines
in Fig.~\ref{fig:sigeff-E}.  Here we observe that the above-discussed
energy and jet transverse momentum dependencies are substantially
reduced in the very high energy limit.
 This is also clearly seen in Fig.~\ref{fig:sigeff-pt},
 where we show the $p_{{\rm t}}^{{\rm cut}}$ dependence of the effective
cross section  for the production of a pair of dijets of
 $p_{{\rm t}}^{{\rm jet}}>p_{{\rm t}}^{{\rm cut}}$
 at the Tevatron and the present LHC energies. We compare
again $\sigma_{pp}^{{\rm eff}}$ calculated with all the discussed
contributions taken into account to the one based on the (2v2) mechanism
only, $\sigma_{pp}^{{\rm eff(2v2)}}$.   To investigate
the relative importance of the soft and hard parton splittings, we show in
 Fig.~\ref{fig:sigeff-pt} the effective cross sections calculated using
 the (2v2) and ${\rm (2v1)}_{{\rm s}}$ contributions,
i.e., accounting  for the nonperturbative parton splitting only
(dotted-dashed lines).  Our results nicely demonstrate
how the two  splitting mechanisms complement each
other. While the   contribution of the perturbative  parton splitting
 decreases for $p_{{\rm t}}^{{\rm cut}}\rightarrow Q_{0}$
(the solid and dotted-dashed lines approach each other), the opposite
applies to the respective nonperturbative contribution:
 it is quite small for high $p_{{\rm t}}^{{\rm cut}}$ but gradually increases
with the decrease of the transverse momentum cutoff and provides the dominant
correction to the simple (2v2) picture when approaching the $p_{{\rm t}}^{{\rm cut}}=Q_{0}$
value.
 As a result, the effective cross section
obtained with both contributions taken into account   is characterized
by a relatively weak $p_{t}^{{\rm cut}}$   dependence, as anticipated
previously in Refs.\  \cite{blo14,blo12a} and indicated by experimental data 
\cite{abe97,aba10,aad13,cha14}.

\begin{figure*}[tbh]
\raggedright{}\includegraphics[clip,width=14cm,height=6cm]{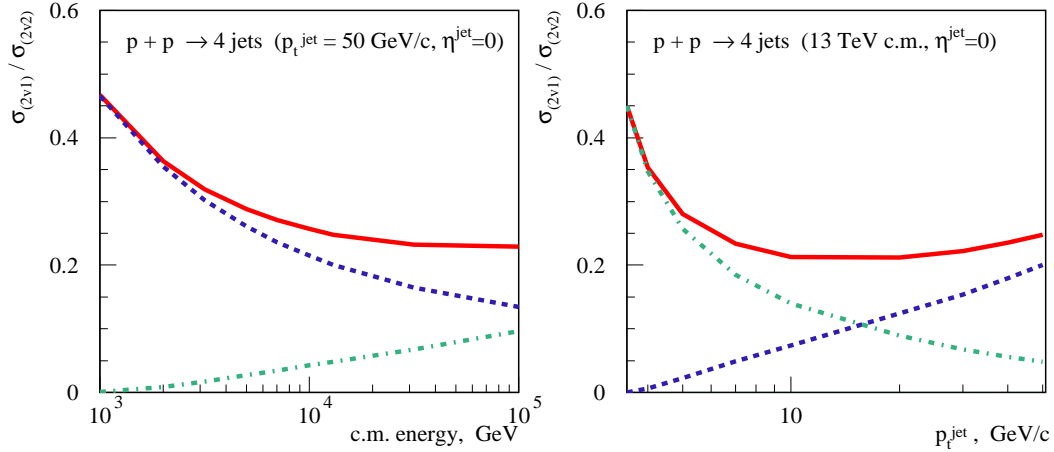}\caption{Energy dependence
 for $p_{{\rm t}}^{{\rm jet}}=50$ GeV/c (left) and
$p_{{\rm t}}^{{\rm jet}}$ dependence for $\sqrt{s}=13$ TeV (right)
of the ratio of the (2v1) to (2v2) contributions to the DPS production of two
dijets of fixed  $p_{{\rm t}}^{{\rm jet}}$ for  $\eta ^{{\rm jet}}=0$
 (solid lines). Partial contributions to this ratio from hard
  and soft   parton splitting mechanisms
are shown by dashed and dotted-dashed lines, respectively.\label{fig:Reff=pt}}
\end{figure*}

 To investigate further the relative roles of the  soft  and  hard 
parton splitting mechanisms, we plot in Fig.~\ref{fig:Reff}
the calculated $p_{{\rm t}}^{{\rm cut}}$ dependence (at $\sqrt{s}=13$
TeV) and the energy dependence (for $p_{{\rm t}}^{{\rm cut}}=50$ GeV/c)
 of the ratio of the (2v1) to (2v2) contributions to the DPS cross
section for the production of two dijets of 
$p_{{\rm t}}^{{\rm jet}}>p_{{\rm t}}^{{\rm cut}}$, 
$\left[\sigma_{pp}^{{\rm 4jet(2v1)}_{{\rm s}}}
+\sigma_{pp}^{{\rm 4jet(2v1)}_{{\rm h}}}\right]
/\sigma_{pp}^{{\rm 4jet(2v2)}}$. 
The partial contributions to this ratio
from soft, $\sigma_{pp}^{{\rm 4jet(2v1)}_{{\rm s}}}/\sigma_{pp}^{{\rm 4jet(2v2)}}$,
and hard, $\sigma_{pp}^{{\rm 4jet(2v1)}_{{\rm h}}}/\sigma_{pp}^{{\rm 4jet(2v2)}}$,
parton splittings are shown by dotted-dashed and dashed lines respectively.
Additionally, in  Fig.~\ref{fig:Reff=pt}, we plot the energy and transverse %
momentum dependencies of the same quantities for central ($\eta =0$)
 production of   a pair of dijets for fixed $p_{{\rm t}}^{{\rm jet}}$.
The plots in  Fig.~\ref{fig:Reff}~(right) and  Fig.~\ref{fig:Reff=pt}~(right)
clearly demonstrate the increasing importance of the contribution
of the  nonperturbative parton splitting when jet $p_{{\rm t}}$ decreases.

Let us now turn to the  energy dependence of the relative 
contributions of soft and hard parton splittings to the (2v1) mechanism,
shown  in Fig.~\ref{fig:Reff} (left) for the production of a pair of dijets
 of $p_{{\rm t}}^{{\rm jet}}>50$ GeV/c, integrated over the phase space,
 and  in Fig.~\ref{fig:Reff=pt} (left) for central production of two dijets
 of fixed  $p_{{\rm t}}^{{\rm jet}}=50$ GeV/c.
The contribution of the nonperturbative parton splitting, plotted by 
dotted-dashed
lines, exhibits a noticeable energy rise,  which is mostly due to the
 increasing rapidity range available for parton evolution.
   In contrast, the relative contribution of the hard splitting 
is gradually reduced with increasing energy. This is mainly because
the  contribution of the perturbative parton splitting to $_{2}{\rm GPDs}$ 
of gluons is effectively one $\ln x$ short in the low $x$ limit, compared to 
the case of two independent parton cascades,\footnote{In the case of the hard
parton splitting, there is a single convolution with the gluon PDF at the 
initial scale $Q_0^2$.}
  as discussed previously in
Refs.\ \cite{blo14,blo12a} (see, e.g.,  Fig.\ 2 in Ref.\  \cite{blo12a}).
Apart from that, an additional reduction arises due to   color fluctuations in 
the interacting protons and    partons' transverse diffusion.

\begin{figure}[tbh]
\centering{}\includegraphics[clip,width=7cm,height=6cm]{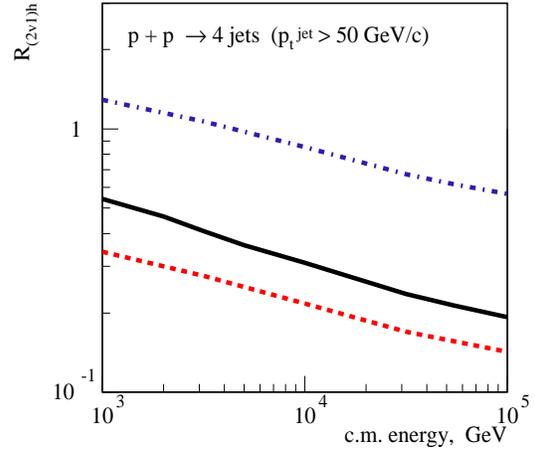}\caption{Energy
dependence  of the  ratio $R_{{\rm (2v1)}_{{\rm h}}}$ of the
  ${\rm (2v1)}_{{\rm h}}$ to (2v2) contributions to the DPS production
  of two dijets of  $p_{{\rm t}}^{{\rm jet}}>50$ GeV/c
  (solid line) and the ratios $R^{(ii)}_{{\rm (2v1)}_{{\rm h}}}$
  for the combinations of large ($i=1$) and small  ($i=2$)  size 
  parton Fock states - dashed and dotted-dashed lines, respectively.
\label{fig:part-Reff}}
\end{figure}

\begin{figure}[tbh]
\centering{}\includegraphics[clip,width=7cm,height=6cm]{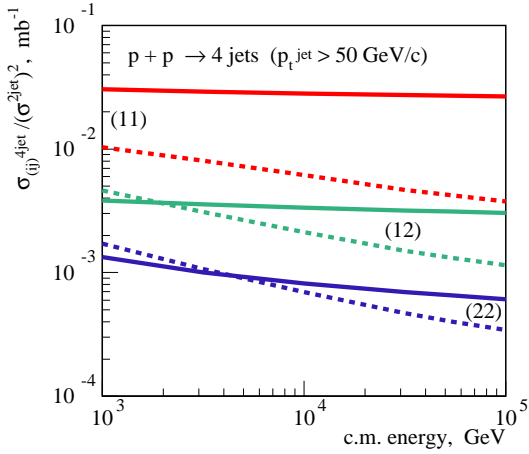}\caption{Energy
dependence  of the  ratios  $2\sigma_{(ij)}^{{\rm 4jet(2v2)}}/
\left(\sigma_{pp}^{2{\rm jet}}\right)^{2}$  (solid lines) and
 $2\sigma_{(ij)}^{{\rm 4jet(2v1)}_{{\rm h}}}/
\left(\sigma_{pp}^{2{\rm jet}}\right)^{2}$ (dashed lines)
for DPS production  of two
 dijets of  $p_{{\rm t}}^{{\rm jet}}>50$ GeV/c
for different combinations of parton Fock states $(ij)$, as indicated in the
figure.
\label{fig:part-sigeff}}
\end{figure}

To elucidate this latter point, let us compare in  Fig.~\ref{fig:part-Reff}
 the energy dependence of the ratios
\begin{equation}
R^{(ii)}_{{\rm (2v1)}_{{\rm h}}}=
\sigma_{(ii)}^{{\rm 4jet(2v1)}_{{\rm h}}}/\sigma_{(ii)}^{{\rm 4jet(2v2)}}
 \label{Rii}
\end{equation}
of the ${\rm (2v1)}_{{\rm h}}$ to (2v2) contributions for 
combinations of large ($i=1$) and small  ($i=2$) size diffractive eigenstates
of the  projectile and target protons, 
where $\sigma_{(ii)}^{{\rm 4jet(2v2)}}$
and $\sigma_{(ii)}^{{\rm 4jet(2v1)}_{{\rm h}}}$ are given by the
corresponding terms (for fixed $i=j$) in the rhs of Eqs.~(\ref{eq:sigma-2v2-RFT})
 and (\ref{eq:sig-DPS-1v2-rft}) respectively. Obviously,
the ratio $R^{(ii)}_{{\rm (2v1)}_{{\rm h}}}$ is much larger in the
case of  the small size parton configurations  ($i=2$), where the contribution
of the perturbative splitting of, e.g., a projectile parton is strongly
enhanced by the large spatial parton density in the target (or vice versa).
The effects of parton diffusion are largely canceled\footnote{In fact,
such a cancellation is precise for  central ($\eta =0$)
 production of   a dijet pair  of fixed $p_{{\rm t}}^{{\rm jet}}$.}
between the numerator and the denominator in Eq.\ (\ref{Rii}), which results
in a very similar energy dependence for $R^{(11)}_{{\rm (2v1)}_{{\rm h}}}$
and  $R^{(22)}_{{\rm (2v1)}_{{\rm h}}}$. In contrast, the ratio of the total
 ${\rm (2v1)}_{{\rm h}}$ to (2v2) contributions,
\begin{equation}
R_{{\rm (2v1)}_{{\rm h}}}=
\left[\sum_{i,j} \sigma_{(ij)}^{{\rm 4jet(2v1)}_{{\rm h}}}\right]
/\left[\sum_{i,j} \sigma_{(ij)}^{{\rm 4jet(2v2)}}\right],
 \label{Rtot}
\end{equation}
 shown by the solid line in  Fig.~\ref{fig:part-Reff}, falls down faster
 with increasing energy.
   This is because 
the effects of parton diffusion are particularly
important for the small-size Fock states, leading to a fast increase
of the effective radius of low-$x$ parton clouds (compared to the
spatial distribution
of large-$x$ partons) and to a quick energy decrease of 
the  respective contributions,
 $\sigma_{(22)}^{{\rm 4jet(2v2)}}$
and $\sigma_{(22)}^{{\rm 4jet(2v1)}_{{\rm h}}}$ (and similarly for the
off-diagonal contributions, $\sigma_{(12)}^{{\rm 4jet(2v2)}}$
and $\sigma_{(12)}^{{\rm 4jet(2v1)}_{{\rm h}}}$).
We illustrate the latter point by comparing in  Fig.~\ref{fig:part-sigeff}
 the ratios $2\sigma_{(ij)}^{{\rm 4jet(2v2)}}/
\left(\sigma_{pp}^{2{\rm jet}}\right)^{2}$ and
 $2\sigma_{(ij)}^{{\rm 4jet(2v1)}_{{\rm h}}}/
\left(\sigma_{pp}^{2{\rm jet}}\right)^{2}$ for different combinations
of the large- and small-size diffractive eigenstates 
as a function of $\sqrt{s}$. Thus, the quicker decrease of 
 the ratio of the total
 ${\rm (2v1)}_{{\rm h}}$ to (2v2) contributions,
$R_{{\rm (2v1)}_{{\rm h}}}$, in  Fig.~\ref{fig:part-Reff} is explained
by the fact that with increasing energy the interactions involving
 small-size parton Fock states contribute less   to the DPS
cross sections $\sigma_{pp}^{4{\rm jet}(\alpha)}(s,p_{{\rm t}}^{{\rm cut}})$
[$\alpha={\rm (2v2)}, {\rm (2v1)}_{{\rm s}}, {\rm (2v1)}_{{\rm h}}$],
compared to the case when both interacting protons are represented by
their largest size Fock states.

It is noteworthy that there is a number of other effects which contribute
to the uncertainty of the hard parton splitting contribution and which
have led to the much smaller values of $R_{{\rm (2v1)}_{{\rm h}}}$,
obtained in this work, compared to previous calculations in 
 Refs.\  \cite{blo14,gau14}. For example, our choice of the factorization scale,
 $M_{{\rm F}}^{2}=p_t^2/4$, shortens somewhat the kinematic range available
 for the perturbative parton evolution, compared to the standard choice
 ($M_{{\rm F}}^{2}=p_t^2$), which reduces the respective collinear enhancements.
 The corresponding uncertainty is related to higher-order effects of which the
 potential importance has already been stressed in  Ref.\  \cite{blo14}.
 Another effect is the sensitivity of  $R_{{\rm (2v1)}_{{\rm h}}}$ to the
 chosen functional form for the two-gluon form factor. For example,
 replacing the dipole ansatz used in Ref.\  \cite{blo14} by a Gaussian reduces
  $R_{{\rm (2v1)}_{{\rm h}}}$ by $\simeq 17$\% [see   Eqs.\ (19-21)
  in that reference].
  Additionally, the relative importance of the perturbative parton splitting
  for DPS  depends noticeably on both the large and small $x$ behavior
  of gluon PDF, $G(x,Q_0^2)$. A steeper low-$x$ rise of  $G(x,Q_0^2)$
  would enhance the (2v2) contribution, compared to ${\rm (2v1)}_{{\rm h}}$,
   and thus reduce  $R_{{\rm (2v1)}_{{\rm h}}}$ (see Eq.\ (5) of 
    Ref.\  \cite{blo14}). On the other hand, a harder large-$x$ shape of the
    gluon PDF would, on the contrary, enhance the hard splitting contribution,
    while leaving the  (2v2) contribution practically unchanged.
       
    Finally,  we have to mention that the obtained values of 
     $\sigma_{pp}^{{\rm eff}}$ are systematically higher than the measured
     values, though   consistent with them within the
     reported  experimental accuracies. Clearly, the above-discussed
     uncertainties concerning the contribution of the perturbative parton
     splitting give one enough freedom to adjust 
      $R_{{\rm (2v1)}_{{\rm h}}}$  and thereby to approach experimental
      measurements. However, the nonperturbative effects studied in this work
      are also subject to substantial uncertainties related, e.g.\ 
      to the strength of the triple-Pomeron coupling, which controls the
      magnitude of the soft parton splitting contribution. On the other hand,
      the employed Good-Walker--like scheme with only two diffractive
      eigenstates
      is rather crude, and a more advanced treatment is generally desirable.
      Hence, we may only speak here about a semiquantitative approach
      to the problem.

\section{Conclusions\label{sec:Outlook}}

In this work, we applied the phenomenological Reggeon field theory
framework to investigate the relative importance of perturbative and
nonperturbative multiparton correlations for the treatment of double
parton scattering in proton-proton collisions. We obtained a significant
correction to the effective cross section for DPS due to nonperturbative
parton splitting. When combined with the contribution of perturbative
parton splitting, this results in a rather weak energy and transverse
momentum dependence of $\sigma_{pp}^{{\rm eff}}$, in agreement with
experimental observations. On the other hand, we observed that color
fluctuations have a noticeable impact on the calculated rates of double
parton scattering and on  the relative role
of the perturbative parton splitting mechanism.

While the obtained numerical results bear a substantial model dependence,
the observed qualitative trends are of general character and can probably
be reproduced applying other alternative approaches to the problem,
for example, using the original color fluctuations framework of
Ref.~\cite{fra08}.

\subsection*{Acknowledgments}

The authors acknowledge useful discussions with B.\ Blok and M.~Strikman. 
This work was supported in part by Deutsche Forschungsgemeinschaft (Project
No.\ OS 481/1-1) and  the State of Hesse via the LOEWE-Center HIC for FAIR.

\end{document}